\documentclass[english,conference,letterpaper, 10pt]{IEEEtran}
\usepackage[T1]{fontenc}
\usepackage[latin9]{inputenc}

\usepackage{geometry}
\usepackage{array}
\usepackage{booktabs}
\usepackage{dsfont}
\usepackage{amsmath}
\usepackage{amssymb}
\usepackage{graphicx}
\usepackage{amsthm}
\usepackage{cite}

\usepackage{textcomp}
\usepackage{xcolor}
\usepackage{babel}

\begin{document}
	
	\title{Robust Semantic Transmission for Low-Altitude UAVs: Predictive Channel-Aware Scheduling and Generative Reconstruction}
	
	\author{
		\IEEEauthorblockN{Jijia Tian, Junting Chen, Pooi-Yuen Kam}
		\IEEEauthorblockA{School of Science and Engineering and Future Network Intelligence Institute (FNii)\\
			The Chinese University of Hong Kong, Shenzhen, Guangdong 518172, China}
	}
	
	\maketitle
	
\begin{abstract}
Unmanned aerial vehicle (UAV) downlink transmission facilitates critical time-sensitive visual applications but is fundamentally constrained by bandwidth scarcity and dynamic channel impairments. The rapid fluctuation of the air-to-ground (A2G) link creates a regime where reliable transmission slots are intermittent and future channel quality can only be predicted with uncertainty. Conventional deep joint source-channel coding (DeepJSCC) methods transmit coupled feature streams, causing global reconstruction failure when specific time slots experience deep fading. Decoupling semantic content into a deterministic structure component and a stochastic texture component enables differentiated error protection strategies aligned with channel reliability. A predictive transmission framework is developed that utilizes a split-stream variational codec and a channel-aware scheduler to prioritize the delivery of structural layout over reliable slots. Experimental evaluations indicate that this approach achieves a 5.6 dB gain in peak signal-to-noise (SNR) ratio over single-stream baselines and maintains structural fidelity under significant prediction mismatch.
\end{abstract}
\section{Introduction}
	\label{sec:introduction}
	UAVs have emerged as a key enabler for future wireless networks, offering flexible deployment for applications such as aerial surveillance, disaster relief, and remote sensing \cite{wu2021uav_overview,liaJiaZha:J26,li2025joint}. Unlike static terrestrial infrastructure, low-altitude UAVs operate in complex three-dimensional environments where the A2G channel is heavily governed by dynamic geometric relationships. While high mobility can facilitate Line-of-Sight (LOS) connectivity, it simultaneously exposes the link to severe propagation challenges. Specifically, the low-altitude regime is susceptible to frequent physical blockages caused by urban obstacles or terrain, resulting in rapid and stochastic transitions between LOS and Non-Line-of-Sight (NLOS) states \cite{alhourani2014modeling}. Coupled with distance-dependent path loss and large-scale shadowing, these factors create a highly volatile channel characterized by deep fades and intermittent outages. Consequently, maintaining high-fidelity image transmission in such fluctuating environments remains a critical challenge, as conventional fixed-rate transmission schemes often fail to adapt to the abrupt variations in instantaneous channel capacity.
	
	To address transmission in such dynamic regimes, the classical separation principle is often insufficient due to the cliff effect, where reconstruction fails catastrophically once the channel capacity falls below the design rate. Recently, DeepJSCC \cite{bourtsoulatze2019icassp} has demonstrated significant potential by mapping source data directly to continuous channel symbols, thereby achieving graceful degradation with decreasing SNR. However, most existing DeepJSCC frameworks rely solely on instantaneous Channel State Information (CSI) or average statistics, ignoring the temporal correlation inherent in UAV trajectories \cite{kurka2020deepjsccf, zhou2025feature, zhang2023predictive, ying2026token}. They operate in a reactive manner, failing to exploit the predictability of channel blockage patterns to optimize resource allocation over a time horizon. Furthermore, standard DeepJSCC schemes typically treat the learned latent representations as a uniform data stream. They lack the semantic granularity to distinguish between critical geometric structures and high-frequency textures, resulting in inefficient bandwidth utilization where essential semantic content is not explicitly prioritized during deep fades.
	
	From a semantic and perceptual perspective, minimizing pixel-level distortion is not always the optimal strategy under strict bandwidth constraints. Recent advances in the Rate-Distortion-Perception (RDP) trade-off \cite{liang2025synonymous} suggest that high perceptual quality can be maintained even with lossy reconstruction, provided that the semantic content is preserved. In natural images, visual information can be conceptually disentangled into a deterministic \textit{structure} component (e.g., edges, shapes, and object layouts) and a stochastic \textit{texture} component (e.g., fine-grained surface details). While structure is essential for correct semantic interpretation and must be accurately transmitted, texture exhibits high statistical redundancy and can often be plausibly synthesized by generative models at the receiver. Existing DeepJSCC approaches, however, entangle these features in a shared latent space. Consequently, when channel quality drops, both structure and texture degrade simultaneously, leading to blurring and semantic artifacts that compromise the utility of the received image.
	
	To address the hostile and fluctuating channel conditions in low-altitude UAV downlinks, we propose a predictive semantic transmission framework that integrates trajectory-driven SNR forecasting with hierarchical feature coding. Unlike existing reactive schemes, our approach utilizes predicted channel states to guide a proactive resource allocation strategy over a finite time horizon. At the core of the system is a Structure-Texture Variational Autoencoder (ST-VAE) that explicitly disentangles image features into a deterministic structural stream and a stochastic texture stream. This decoupling enables a channel-aware predictive scheduler to prioritize the transmission of essential geometric skeletons while opportunistically scheduling texture blocks based on the forecasted bandwidth. For slots affected by outages or restricted budgets, the receiver employs a conditional generative prior to hallucinate missing textures, thereby maintaining high perceptual fidelity. Extensive performance evaluations demonstrate that the proposed method achieves a substantial 5.6 dB gain in PSNR compared to standard DeepJSCC and remains remarkably robust under significant channel prediction mismatches, validating the effectiveness of combining predictive scheduling with generative semantic reconstruction.

\section{System Model}
\label{sec:system_model}

We consider a low-altitude UAV-assisted downlink system where a single UAV transmits semantic image data to a ground user (GU). 
The system operates over a transmission horizon discretized into $K$ time slots, each of duration $T$. 
The channel statistics are assumed to be quasi-static within each slot but may vary across slots due to dynamic geometry changes and large-scale fading.

\subsection{Geometric Configuration and Trajectory}
The GU is located at a fixed 3D coordinate $\mathbf{w} = [x_g, y_g, 0]^T \in \mathbb{R}^3$. At time slot $k$, the UAV position is $\mathbf{q}_k = [x_u(k), y_u(k), h_u(k)]^T \in \mathbb{R}^3$. The instantaneous UAV-GU distance is given by:
\begin{equation}
\begin{aligned}
    d_k &= \|\mathbf{q}_k - \mathbf{w}\|_2 \\
    &= \sqrt{(x_u(k) - x_g)^2 + (y_u(k) - y_g)^2 + h_u(k)^2}.
\end{aligned}
\end{equation}
The elevation angle $\theta_k$ is defined as:
\begin{equation}
    \theta_k = \arcsin\left( \frac{h_u(k)}{d_k} \right)
\end{equation}
which dictates the LOS probability in low-altitude propagation environments.

\subsection{Probabilistic A2G Channel Model}
The large-scale attenuation consists of distance-dependent path loss conditioned on the LOS/NLOS state and a correlated shadowing process.

\subsubsection{LOS/NLOS State}
Let $o_k \in \{\text{L}, \text{N}\}$ denote the LOS and NLOS states, respectively. The LOS probability is modeled as a logistic function of the elevation angle:
\begin{equation}
    P_{\text{LOS}}(\theta_k) = \frac{1}{1 + \alpha_{\text{LOS}} \exp(-\beta_{\text{LOS}} (\theta_k - \theta_0))}
\end{equation}
where $\alpha_{\text{LOS}}$, $\beta_{\text{LOS}}$, and $\theta_0$ are environment-dependent constants. 
In this work, we adopt a simplified threshold-based state model for tractability:
\begin{equation}
    o_k = \begin{cases} 
    \text{L}, & \theta_k \geq \theta_{\text{th}}, \\
    \text{N}, & \text{otherwise}
    \end{cases}
\end{equation}
where $\theta_{\text{th}}$ is the environment-specific elevation threshold.
\subsubsection{Path Loss and Shadowing}
The path loss in dB for state $o_k$ is:
\begin{equation}
    PL_{o_k}(d_k) = 20 \log_{10}\left( \frac{4\pi f_c d_k}{c} \right) + \eta_{o_k}
\end{equation}
where $f_c$ is the carrier frequency, $c$ is the speed of light, and $\eta_{o_k}$ represents the excessive loss (typically $\eta_{\text{N}} > \eta_{\text{L}}$). 
To capture temporal correlation, the shadowing component $\chi_k$ is modeled as a first-order autoregressive (AR(1)) process:
\begin{equation}
    \chi_k = \rho(v_k) \chi_{k-1} + \sqrt{1 - \rho(v_k)^2} \, \xi_k
\end{equation}
where $\xi_k \sim \mathcal{N}(0, \sigma_{o_k}^2)$ is the innovation noise, and the correlation coefficient is $\rho(v_k) = \exp(-v_k T / d_{\text{corr}})$, with $v_k$ being the UAV speed and $d_{\text{corr}}$ the decorrelation distance. 
The total large-scale attenuation in dB is denoted as $L^{\text{dB}}_k = PL_{o_k}(d_k) + \chi_k$.

\subsection{Signal Transmission Model}
\begin{figure}[!t]
\centering
\includegraphics[width=1\linewidth]{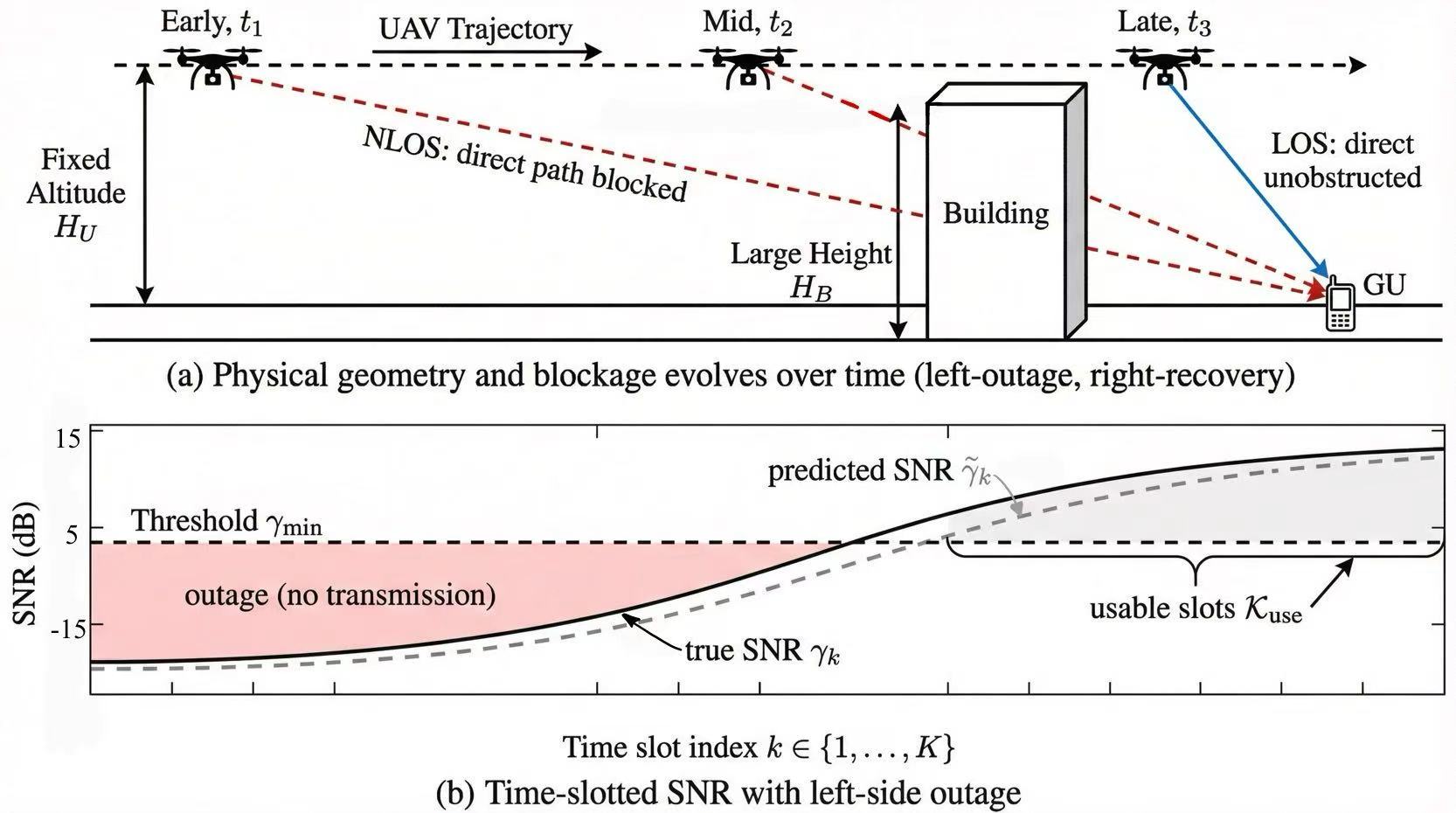}
\caption{Illustration of the time-varying UAV downlink induced by geometry and blockage. 
(a) The UAV trajectory and an obstacle lead to LOS/NLOS transitions. 
(b) The resulting time-slotted SNR sequence exhibits an outage interval where $\gamma_k < \gamma_{\min}$ (no transmission allowed) and a usable-slot set $\mathcal{K}_{\text{use}}$ for scheduling based on SNR prediction.}
\label{fig:scenario}
\end{figure}

We adopt an analog DeepJSCC strategy where real-valued latent features are mapped directly to channel inputs without explicit quantization. 
After coherent phase compensation, the equivalent baseband channel in slot $k$ is modeled as a real scalar gain. 
Let $s \in \mathbb{R}$ denote a transmitted symbol within slot $k$, subject to the average power constraint $\mathbb{E}\{|s|^2\} \le P_t$. 
The received signal is:
\begin{equation}
    y = g_k s + n
\end{equation}
where $n \sim \mathcal{N}(0, N_0 B)$ is the additive white Gaussian noise (AWGN), with $N_0$ denoting the noise power spectral density and $B$ representing the system bandwidth. The channel power gain is defined by $g_k^2 \triangleq 10^{-L^{\text{dB}}_k/10}$. The instantaneous SNR for slot $k$ is given by:
\begin{equation}
    \gamma_k = \frac{P_t}{N_0 B} 10^{-L^{\text{dB}}_k/10}.
\end{equation}
A slot is considered \textit{usable} for semantic transmission only if its SNR exceeds a minimum threshold $\gamma_{\min}$. We define the binary usability indicator as:
\begin{equation}
    a_k = \mathds{1}(\gamma_k \geq \gamma_{\min})
\end{equation}
where $\mathds{1}\{\cdot\}$ is the indicator function. This mechanism translates the physical blockage and fading dynamics into a discrete usable-slot set $\mathcal{K}_{\text{use}}$ for higher-layer scheduling. Fig.~\ref{fig:scenario} illustrates how blockage and LOS/NLOS transitions along the UAV trajectory translate into a time-slotted SNR sequence with outage intervals.

\subsection{Data-Driven SNR Prediction}
Effective scheduling relies on predicting channel quality over the forthcoming transmission horizon. A neural predictor $F_{\phi}$ estimates future SNRs based on historical measurements and the deterministic planned trajectory. Let $M$ denote the observation history length and let $K$ denote the prediction horizon. At slot $k$, the input state consists of the historical SNR sequence $\mathbf{H}_{\mathrm{snr}}^{(k)} \in \mathbb{R}^M$, the historical trajectory $\mathbf{H}_{\mathrm{traj}}^{(k)} \in \mathbb{R}^{M \times 3}$, and the planned future trajectory $\mathbf{P}_{\mathrm{plan}}^{(k)} \in \mathbb{R}^{K \times 3}$, where $\mathbf{P}_{\mathrm{plan}}^{(k)} = [\mathbf{q}_{k+1}, \dots, \mathbf{q}_{k+K}]$. The input feature vector is aggregated as $\mathbf{u}_k = (\mathbf{H}_{\mathrm{snr}}^{(k)}, \mathbf{H}_{\mathrm{traj}}^{(k)}, \mathbf{P}_{\mathrm{plan}}^{(k)})$. The predicted SNR sequence is obtained via
\begin{equation}
    [\hat{\gamma}_{k+1}, \dots, \hat{\gamma}_{k+K}] = F_{\phi}(\mathbf{u}_k).
\end{equation}
These predictions determine the budget allocation and block scheduling in Section~\ref{sec:realization}, while the physical transmission is subject to the realized SNR $\gamma_k$.

	
\section{Semantic Rate Distortion Perception Formulation}
\label{sec:formulation}

We consider semantic image transmission over a time slotted UAV downlink with realized SNR sequence $\{\gamma_k\}_{k=1}^{K}$ and predicted SNR sequence $\{\hat{\gamma}_k\}_{k=1}^{K}$. Let $X\sim p(x)$ denote the source image and $\hat{X}$ denote the reconstruction at the receiver.

Communication cost is measured by the number of real valued baseband samples that can be transmitted within each slot. Under a fixed system bandwidth and a fixed slot duration, the transmitter can inject only a finite length baseband sample sequence into the channel during slot $k$. Let $n_k\in\mathbb{Z}_{+}$ denote this per slot sample budget and assume $\sum_{k=1}^{K} n_k = n_{\mathrm{tot}}$ for a fixed per image total budget. The realized SNR sequence governs the corruption level of the transmitted samples, while the predicted SNR sequence is used only to allocate the sample budgets across slots and to decide which semantic blocks are placed into which slots.

A policy $\pi$ specifies an encoder, a block formation rule, a block to slot scheduler and a decoder. Given a realization $x$ the encoder produces a latent representation that is partitioned into a finite block set $\mathcal{B}$. Each block $\ell\in\mathcal{B}$ is serialized into $r_{\ell}\in\mathbb{Z}_{+}$ real valued channel symbols. The scheduling decision is represented by binary variables $\{b_{\ell,k}\}$ where $b_{\ell,k}=1$ indicates that block $\ell$ is transmitted in slot $k$. Feasibility under per slot budgets is enforced by
\begin{equation}
\sum_{\ell\in\mathcal{B}} r_{\ell} b_{\ell,k} \le n_k,\quad \forall k.
\label{eq:slot_budget_formulation}
\end{equation}
The channel output is generated according to the realized SNR sequence $\{\gamma_k\}_{k=1}^{K}$. The decoder produces $\hat{x}$ from the received blocks and a completion rule for missing blocks. This end to end mapping induces a conditional reconstruction distribution $p_{\pi}(\hat{x}\mid x)$ due to channel noise and possible stochastic completion.

\subsection{Synonymity Based Reconstruction Model}
\label{subsec:synset}

For a given realization $x$ the admissible reconstruction set is approximated by a proxy criterion that combines a distortion metric and a perceptual discrepancy metric~\cite{liang2025synonymous}. This proxy set represents reconstructions that are acceptable under the operating fidelity and perceptual requirements. Let $d(x,\hat{x})$ denote a distortion metric and let $d_p(p(x),p_{\pi}(\hat{x}))$ denote a perceptual discrepancy proxy. The induced expected distortion is
\begin{equation}
D(\pi) \triangleq \mathbb{E}_{x \sim p(x)}\,\mathbb{E}_{\hat{x}\sim p_{\pi}(\hat{x}\mid x)} \left[d(x,\hat{x})\right].
\label{eq:D_def}
\end{equation}
The induced perceptual discrepancy is
\begin{equation}
P(\pi) \triangleq d_p(p(x),p_{\pi}(\hat{x}))
\label{eq:P_def}
\end{equation}
where $p_{\pi}(\hat{x})$ is the marginal distribution induced by $p(x)$ and $p_{\pi}(\hat{x}\mid x)$.

\subsection{Abstract Objective Under Channel Use Budgets}
\label{subsec:rdp}

Classically the rate distortion perception tradeoff can be written in terms of mutual information and distortion perception constraints. In the slotted UAV downlink setting the rate is replaced by hard channel use budgets with per slot constraints induced by $\{n_k\}_{k=1}^{K}$. A feasible policy must satisfy the budget constraints through its block to slot allocation rule. Let $\Pi\{n_1,\ldots,n_K\}$ denote the set of policies whose induced allocations satisfy \ref{eq:slot_budget_formulation} and the per block single assignment constraint in Section~\ref{sec:realization}.

The semantic transmission objective is modeled as
\begin{align}
\min_{\pi} \quad & D(\pi) \nonumber\\
\text{s.t.}\quad & P(\pi) \le P_0, \label{eq:abstract_problem}\\
& \pi \in \Pi\{n_1,\ldots,n_K\}. \nonumber
\end{align}
Section~\ref{sec:realization} provides a parametric neural realization that enforces feasibility by construction and optimizes a Lagrangian surrogate of (\ref{eq:abstract_problem}).

\section{Neural Realization ST-VAE and Predictive Scheduling}
\label{sec:realization}

\begin{figure}[t]
\centering
\includegraphics[width=1\linewidth]{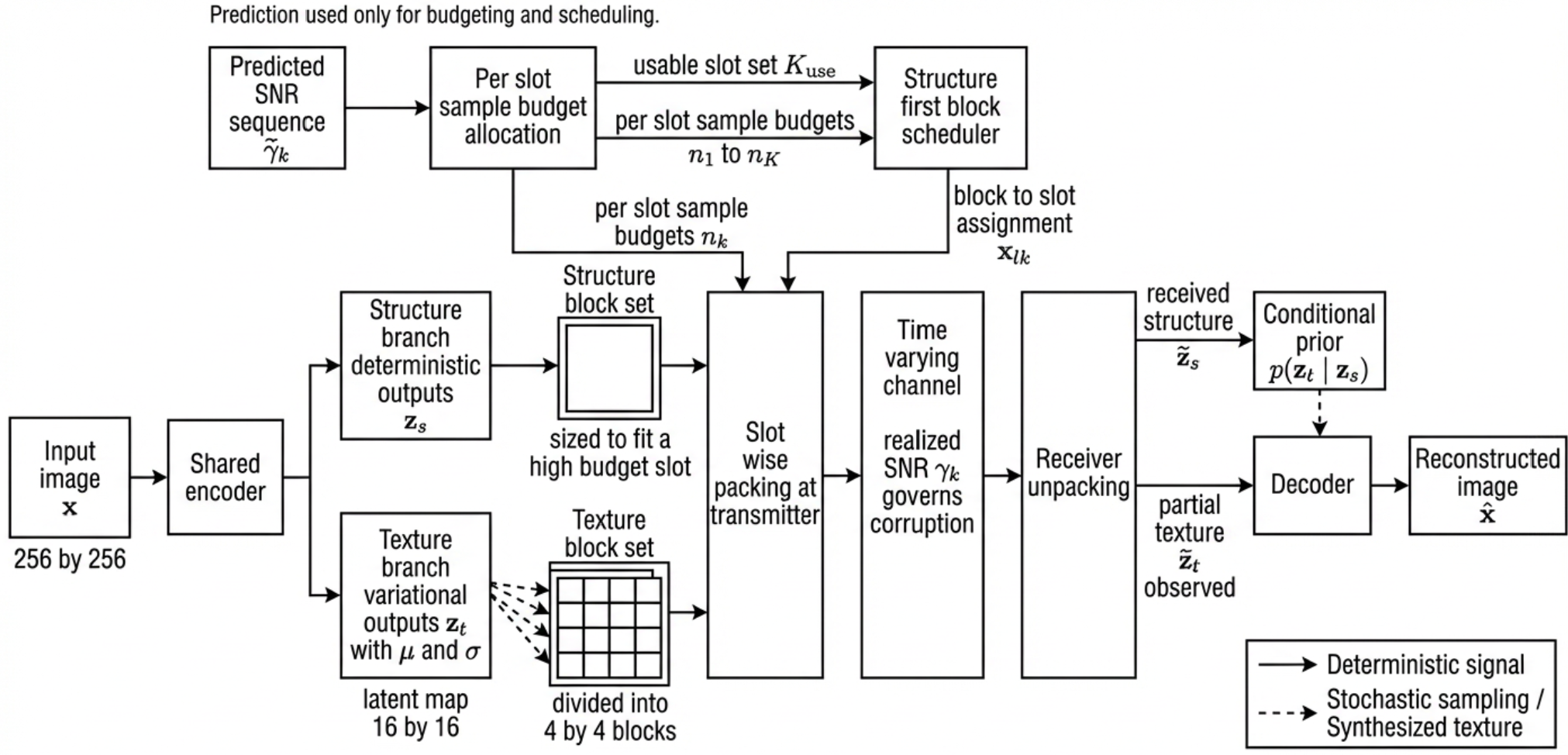}
\caption{Overview of the ST-VAE framework.}
\label{fig:arch}
\end{figure}

This section provides a parametric neural realization of the abstract objective in (\ref{eq:abstract_problem}). The realization consists of an ST-VAE semantic codec and a predictive scheduler shown as Fig. \ref{fig:arch}. The predicted SNR sequence is used to allocate per slot sample budgets and to assign latent blocks to slots. The realized SNR sequence governs channel corruption. Missing texture blocks are completed at the receiver by a conditional prior driven by the received structure.

\subsection{ST-VAE Semantic Codec}
\label{subsec:codec}

The semantic codec operates on a spatial feature map latent to enable block wise transmission. For an input image $x\in\mathbb{R}^{3\times 256\times 256}$, a shared encoder $f_{\mathrm{enc}}$ produces a latent feature map
\begin{equation}
h=f_{\mathrm{enc}}(x)\in\mathbb{R}^{256\times 16\times 16}.
\end{equation}
The structure branch outputs a deterministic structure map
\begin{equation}
z_s=f_s(h)\in\mathbb{R}^{128\times 16\times 16}.
\end{equation}
The texture branch defines a diagonal Gaussian approximate posterior
\begin{equation}
	q_{\phi}(z_t \mid x) \equiv q_{\phi}(z_t \mid h)
	=
	\mathcal{N}\!\left(\mu_{\phi}(h), \mathrm{diag}\!\left[\sigma_{\phi}^2(h)\right]\right)
\end{equation}
where $\mu_{\phi}(h), \log \sigma_{\phi}^2(h) \in \mathbb{R}^{128 \times 16 \times 16}$ are produced by convolutional heads. The texture latent $z_t \in \mathbb{R}^{128 \times 16 \times 16}$ is sampled via the reparameterization trick during training to enable gradient descent.

To enable receiver-side completion, a conditional prior is introduced
\begin{equation}
	p_{\psi}(z_t \mid z_s)
	=
	\mathcal{N}\!\left(\mu_{\psi}(z_s), \mathrm{diag}\!\left[\sigma_{\psi}^2(z_s)\right]\right).
\end{equation}
The conditional prior maps $z_s$ to $\mu_{\psi}(z_s)$ and $\log \sigma_{\psi}^2(z_s)$ using a lightweight convolutional network, and $\log \sigma_{\psi}^2(z_s)$ is bounded for numerical stability.

The decoder reconstructs the image from the received structure and texture tensors
\begin{equation}
\hat{x}=g_{\theta}(\tilde{z}_s,\tilde{z}_t)
\end{equation}
where $\tilde{z}_s$ and $\tilde{z}_t$ are formed from the transmitted blocks after channel corruption and block completion. The decoder mirrors the encoder using transposed convolution layers followed by a final convolution and $\tanh$ mapping to the normalized pixel range.

Block formation is defined on the latent tensors to match the per slot sample budgets. Let $\mathcal{B}_s$ denote the structure block index set and let $\mathcal{B}_t$ denote the texture block index set. Let $\mathcal{B}=\mathcal{B}_s\cup\mathcal{B}_t$. The structure representation is treated as a single global block (which means $|\mathcal{B}_s|=1$) and the texture latent is divided into a $4\times 4$ spatial grid yielding $16$ texture blocks. Each block $\ell\in\mathcal{B}$ is serialized into a vector of $r_{\ell}\in\mathbb{Z}_{+}$ real valued baseband samples.

Given a scheduling decision $\{b_{\ell,k}\}$, the transmitter places the serialized blocks into the corresponding slots, which are corrupted by the realized SNR sequence. At the receiver, the blocks are deserialized and assembled into $\tilde{z}_s$ and a partial texture tensor; missing texture blocks are sampled from $p_{\psi}(z_t\mid z_s)$ conditioned on $\tilde{z}_s$ to form $\tilde{z}_t$.

The abstract objective in (\ref{eq:abstract_problem}) minimizes the expected distortion $D(\pi)$ subject to a perceptual constraint $P(\pi)\le P_0$ under feasibility induced by per-slot budgets. 
In the proposed realization, $D(\pi)$ is approximated by a reconstruction distortion proxy and $P(\pi)$ is controlled by a perceptual proxy together with a conditional generative regularization that governs synthesized texture blocks. 
The reconstruction loss is
\begin{equation}
	\mathcal{L}_{\mathrm{rec}}(x,\hat{x})
	=
	\lambda_{\mathrm{pix}}\|x-\hat{x}\|_2^2
	+
	\lambda_{\mathrm{perc}}\,\ell_{\mathrm{perc}}(x,\hat{x})
	\label{eq:Lrec}
\end{equation}
where $\lambda_{\mathrm{pix}}$ and $\lambda_{\mathrm{perc}}$ are non-negative weighting hyperparameters, and $\ell_{\mathrm{perc}}$ is the perceptual loss.

The conditional prior regularizes the texture posterior
\begin{equation}
\mathcal{L}_{\mathrm{KL},t}
=
D_{\mathrm{KL}}\left[
q_{\phi}(z_t\mid x)\ \Vert\ p_{\psi}(z_t\mid z_s)
\right].
\label{eq:Lkl_t}
\end{equation}
A structure regularizer controls the second moment of the conditioning signal
\begin{equation}
\mathcal{L}_{\mathrm{str}} = \|z_s\|_2^2.
\label{eq:Lstr}
\end{equation}
The training objective is the Lagrangian surrogate
\begin{equation}
\mathcal{L}
=
\mathbb{E}_{x}\left[
\mathcal{L}_{\mathrm{rec}}(x,\hat{x})
+
\beta_t \mathcal{L}_{\mathrm{KL},t}
+
\lambda_{\mathrm{str}}\mathcal{L}_{\mathrm{str}}
\right].
\label{eq:train_obj}
\end{equation}
where $\lambda_{\mathrm{str}}$ and $\beta_t$ are non-negative weighting hyperparameters.

During training, texture blocks are randomly masked and reconstructed via the conditional prior to expose the decoder and the prior to diverse missing patterns. Additive noise is injected into the structure input of the conditional prior to reduce train--test mismatch under channel-corrupted structure.

\subsection{Predictive Scheduler Under Per Slot Sample Budgets}
\label{subsec:scheduler}

Let $\gamma_{\min}$ denote the usability threshold. The predicted SNR sequence defines the usable slot set
\begin{equation}
\mathcal{K}_{\mathrm{use}} \triangleq \left\{ k : \hat{\gamma}_k \ge \gamma_{\min} \right\}.
\end{equation}
Predicted SNR in dB is converted to linear scale by $\hat{\gamma}^{\mathrm{lin}}_k = 10^{\hat{\gamma}^{\mathrm{dB}}_k/10}$. Let $n_{\mathrm{tot}}\in\mathbb{Z}_{+}$ denote the fixed total sample budget per image. Define the slot weight
\begin{equation}
	\bar{c}_k = \mathds{1}\{k \in \mathcal{K}_{\mathrm{use}}\}\log_2\!\left(1+\hat{\gamma}^{\mathrm{lin}}_k\right).
\end{equation}
Here, we choose $\log_2(1+\tilde{\gamma}^{\mathrm{lin}}_k)$ as a monotonic reliability-to-weight mapping inspired by the AWGN capacity expression, and use it purely as a heuristic to rank slots by their effective information-carrying capability under analog transmission. Thus, the proportional allocation is
\begin{equation}
\hat{n}_k = n_{\mathrm{tot}} \frac{\bar{c}_k}{\sum_{j=1}^{K}\bar{c}_j + \epsilon}
\end{equation}
with $\epsilon>0$. Integer per slot budgets are obtained by
\begin{equation}
n_k = \lfloor \hat{n}_k \rfloor.
\end{equation}
Let $R = n_{\mathrm{tot}} - \sum_{k=1}^{K} n_k$ denote the remaining budget. The remainder is assigned by increasing $n_k$ by one for the $R$ indices in $\mathcal{K}_{\mathrm{use}}$ with the largest fractional parts of $\hat{n}_k$, which ensures $\sum_{k=1}^{K} n_k = n_{\mathrm{tot}}$ and yields $n_k=0$ for predicted outage slots.

Scheduling assigns blocks to slots under the per slot budgets. The decision is represented by binary variables $\{b_{\ell,k}\}$ where $b_{\ell,k}=1$ indicates that block $\ell$ is transmitted in slot $k$. The per block single assignment constraint is
\begin{equation}
\sum_{k\in\mathcal{K}_{\mathrm{use}}} b_{\ell,k} \le 1,\quad \forall \ell\in\mathcal{B}.
\label{eq:once}
\end{equation}
The per slot budget constraint is
\begin{equation}
\sum_{\ell\in\mathcal{B}} r_{\ell} b_{\ell,k} \le n_k,\quad \forall k\in\mathcal{K}_{\mathrm{use}}.
\label{eq:slot_budget}
\end{equation}
A structure first rule is enforced by ordering all structure indices before texture indices. Let $\mathcal{L}$ denote the resulting ordered list and let texture indices follow a fixed spatial order. The scheduler greedily fills each usable slot according to the remaining budget.

\section{Simulation Results}
\label{sec:results}

\subsection{Simulation Setup}
\label{subsec:exp_setup}
We utilizes the MS COCO 2017 validation dataset for our experiments. Images are center-cropped to $256\times 256$ pixels with values normalized to $[-1,1]$. The semantic encoder generates a latent map split into a deterministic structure component ($C_{\mathrm{str}}=128$ channels) and a variational texture component ($C_{\mathrm{tex}}=128$ channels). The texture component is partitioned into a $4\times 4$ spatial grid yielding 16 texture blocks. During training, a random texture dropping probability $p_{\mathrm{drop}}=0.3$ forces the receiver to learn conditional completion based on the received structure.

The transmission horizon is $K=10$ slots. A slot is deemed usable if the predicted SNR $\hat{\gamma}_k$ exceeds $\gamma_{\min}=5$ dB. The total bandwidth budget is strictly constrained to $n_{\mathrm{tot}}=512$ real-valued symbols per image, corresponding to a compression ratio of approximately $0.0026$. The channel follows the block-fading model with temporal correlation $\rho=0.9$ described in Section II.

We compare the proposed method against three baselines: 
1) \textit{DeepJSCC}, a single-stream joint source-channel coding scheme with uniform power allocation; 
2) \textit{Uniform Scheduling}, the proposed ST-VAE codec where the total budget is allocated uniformly across slots ($n_k = \lfloor n_{\mathrm{tot}}/K \rfloor$) regardless of channel conditions; and 
3) \textit{No Generation}, the proposed predictive scheduler without the receiver-side generative conditional prior, where missing texture blocks are replaced with zeros.

Reconstruction fidelity is quantified using the Peak Signal-to-Noise Ratio (PSNR) to measure pixel-level distortion and the Structural Similarity Index (SSIM) to capture perceptual structural correlation. All reported metrics are averaged over the evaluation set.

\subsection{Performance versus Average Realized SNR}
\label{subsec:snr_sweep}
\begin{figure}[t]
\centering
\includegraphics[width=0.9\linewidth]{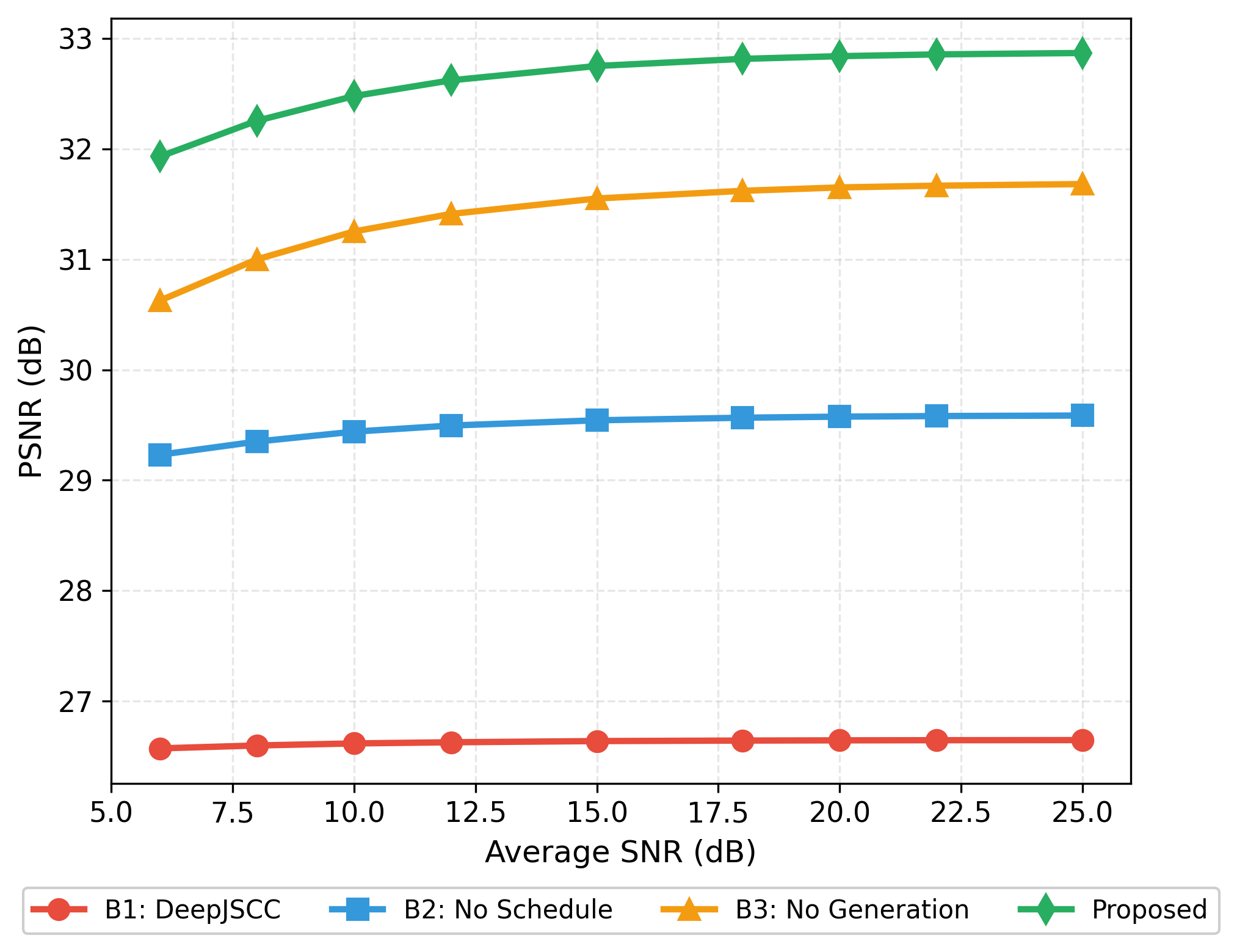}
\caption{Peak signal to noise ratio versus average realized signal to noise ratio.}
\label{fig:snr_psnr}
\end{figure}

\begin{figure}[t]
\centering
\includegraphics[width=0.9\linewidth]{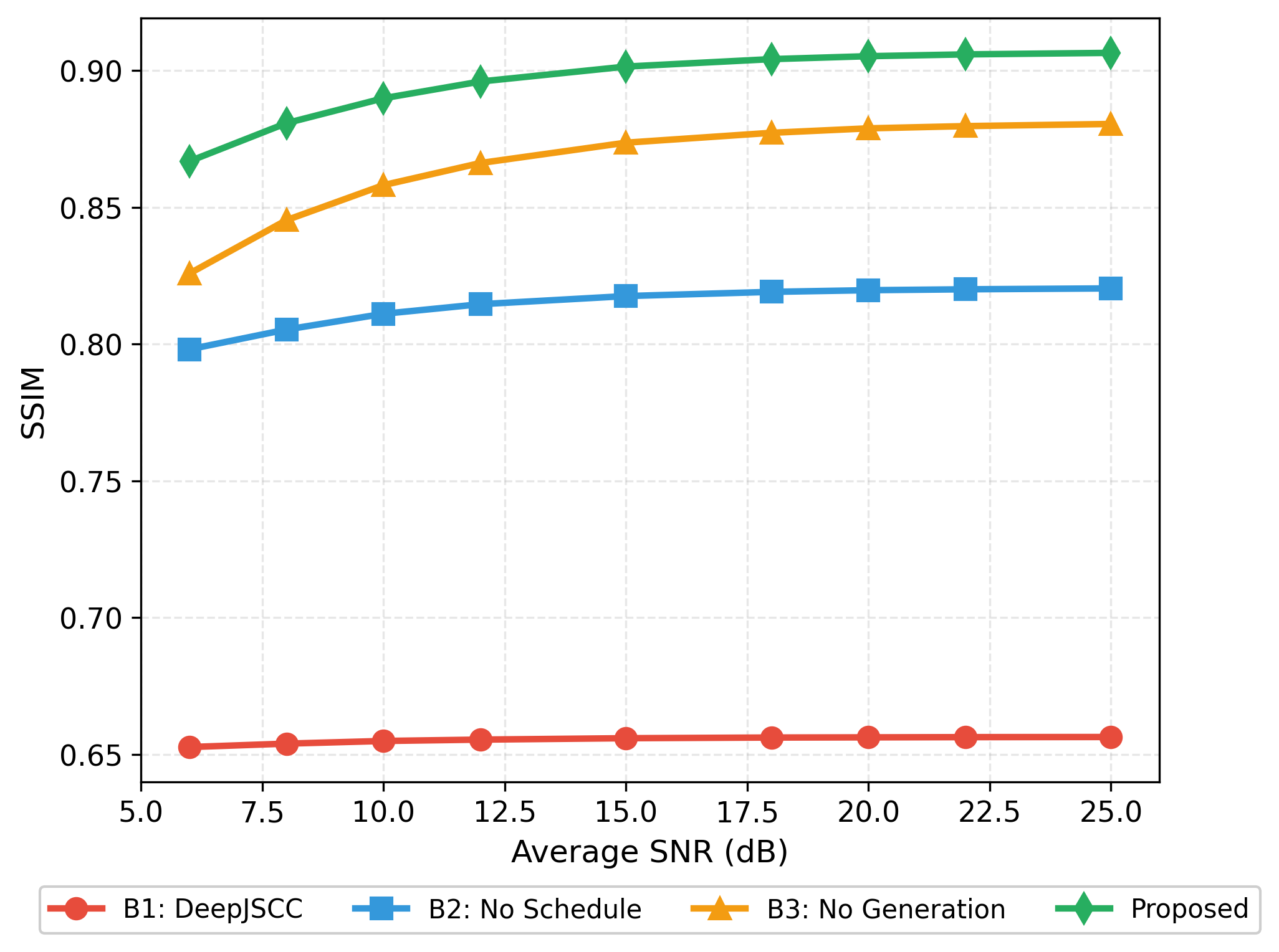}
\caption{Structural similarity index versus average realized signal to noise ratio.}
\label{fig:snr_ssim}
\end{figure} 

Fig.~\ref{fig:snr_psnr} and Fig.~\ref{fig:snr_ssim} present the PSNR and SSIM performance as a function of the average realized SNR. All schemes exhibit monotonic improvement with increasing SNR due to reduced channel corruption. The proposed method consistently outperforms the baselines across the evaluated range of 5 dB to 25 dB. Specifically, at an average SNR of 15 dB, the proposed method achieves a PSNR of approximately 32.8 dB, whereas the DeepJSCC baseline saturates at 27.2 dB. This corresponds to a performance gain of 5.6 dB. Over the entire SNR range, the proposed method maintains a gain between 4.8 dB and 5.8 dB relative to DeepJSCC.

Comparisons with the ablation baselines quantify the contributions of specific modules. The proposed method surpasses the Uniform Scheduling baseline by approximately 3.3 dB at 15 dB SNR, verifying the efficacy of the channel-aware budget allocation algorithm. Furthermore, the proposed method exceeds the No Generation baseline by approximately 1.0 dB, demonstrating that the conditional generative prior effectively enhances visual quality by synthesizing high-frequency texture details when reliable transmission is infeasible.

\subsection{Robustness to Prediction Mismatch}
\label{subsec:sigma_sweep}
\begin{figure}[t]
\centering
\includegraphics[width=0.9\linewidth]{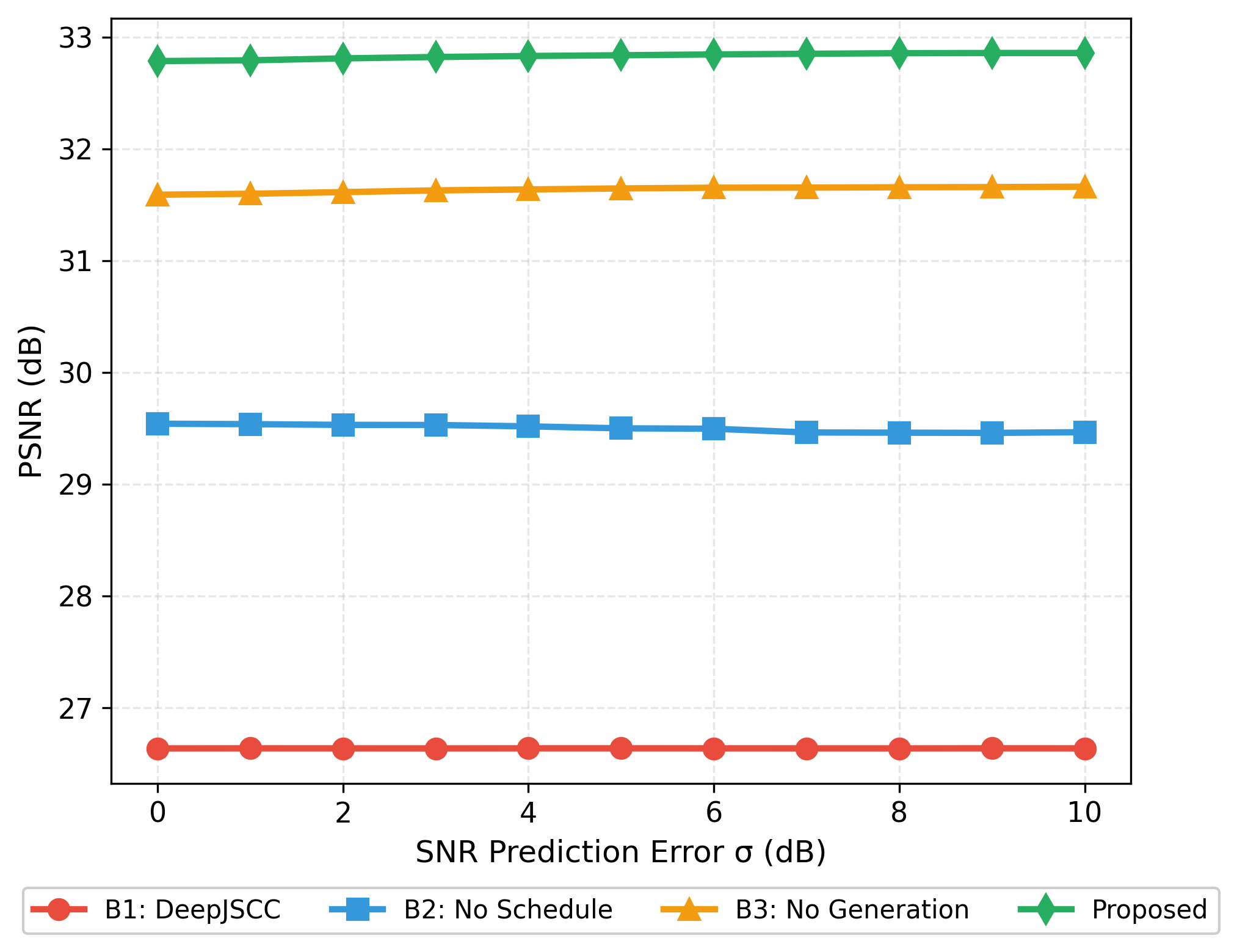}
\caption{Peak signal to noise ratio versus prediction mismatch standard deviation in decibel domain.}
\label{fig:sigma_psnr}
\end{figure}

\begin{figure}[t]
\centering
\includegraphics[width=0.9\linewidth]{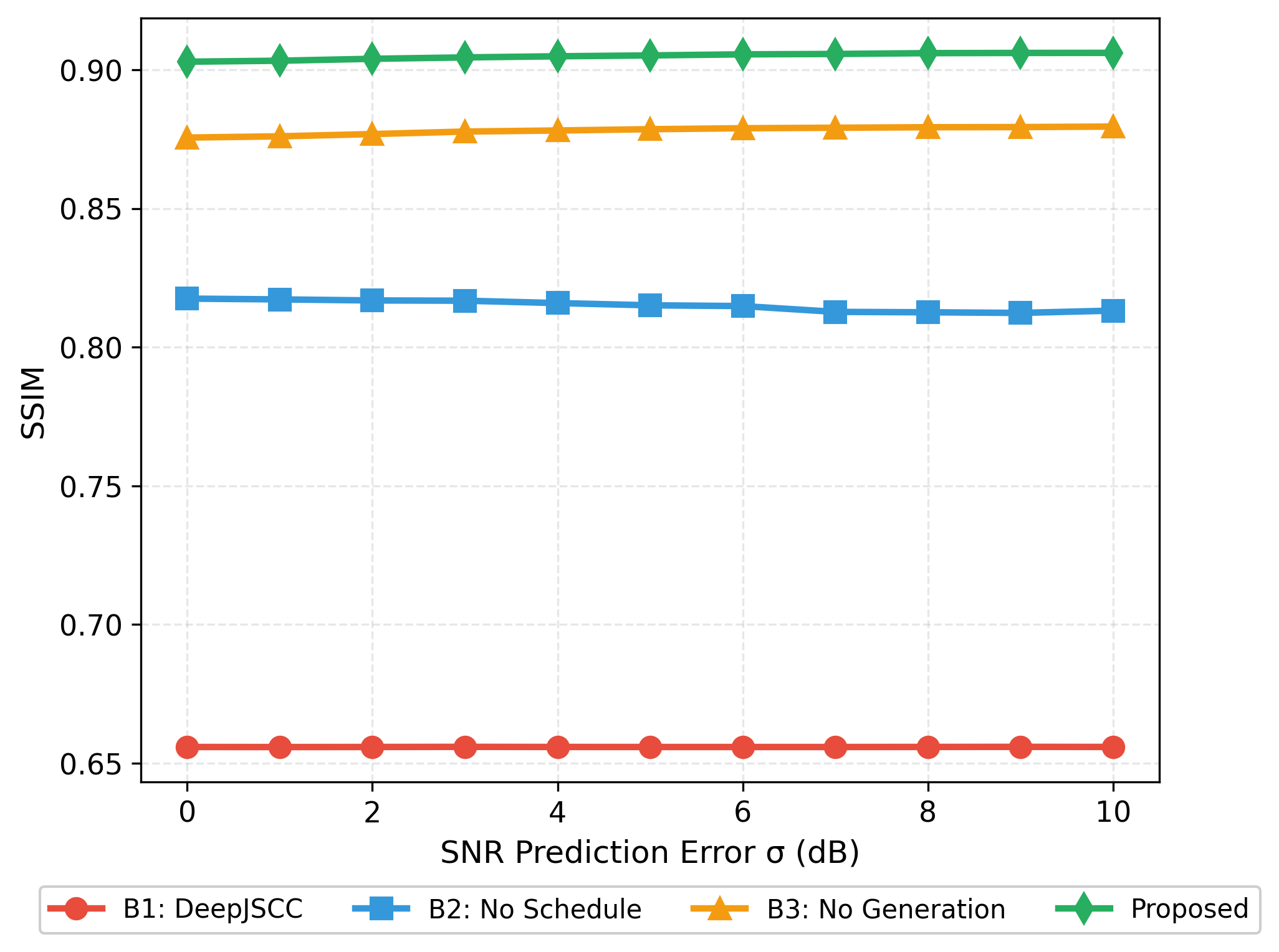}
\caption{Structural similarity index versus prediction mismatch standard deviation in decibel domain.}
\label{fig:sigma_ssim}
\end{figure}

We evaluate robustness by introducing imperfect CSI. The predicted SNR is modeled as $\hat{\gamma}^{\mathrm{dB}}_k = \gamma^{\mathrm{dB}}_k + \epsilon_k$, where $\gamma^{\mathrm{dB}}_k$ is the true channel SNR and $\epsilon_k \sim \mathcal{N}(0, \sigma_{\mathrm{err}}^2)$ represents the prediction error. The average realized SNR is fixed at 15 dB. Fig.~\ref{fig:sigma_psnr} and Fig.~\ref{fig:sigma_ssim} illustrate the impact of the prediction error standard deviation $\sigma_{\mathrm{err}}$ on reconstruction quality.

The proposed method exhibits high resilience to prediction mismatch. As $\sigma_{\mathrm{err}}$ increases from 0 dB to 10 dB, the PSNR of the proposed method remains stable at approximately 32.8 dB, showing negligible degradation. The performance advantage over the DeepJSCC baseline remains constant at 5.6 dB across the uncertainty range. Compared to the Uniform Scheduling baseline, which achieves 31.5 dB, the proposed method maintains a 3.3 dB advantage. This robustness is attributed to the hierarchical decoupling of semantics; the scheduler prioritizes the deterministic structure stream, ensuring geometric fidelity even under suboptimal budget allocation, while the generative prior compensates for texture losses caused by prediction errors.
\section{Conclusion}
This work established a predictive semantic transmission framework for time-slotted UAV downlinks characterized by strict bandwidth constraints and channel uncertainty. By mathematically decoupling the latent representation into a structure block and several texture blocks, the proposed architecture enables differentiated error protection strategies. Numerical evaluations demonstrate that the predictive scheduler, which prioritizes deterministic structure transmission based on estimated channel quality, achieves a coding gain of approximately 5.6 dB in PSNR over DeepJSCC at an average SNR of 15 dB. Furthermore, the integration of a receiver-side conditional generative prior ensures reconstruction stability against channel state information mismatch, maintaining high structural similarity even when prediction errors reach a standard deviation of 10 dB. These results validate that aligning semantic importance with channel reliability effectively mitigates the cliff effect observed in entangled deep communication systems.
\bibliographystyle{IEEEtran}

\end{document}